\documentclass[aps,nofootinbib,showpacs,preprintnumbers,twocolumn,superscriptaddress,prb]{revtex4-1}

\usepackage{amsmath, amsthm, amssymb}
\usepackage{graphicx}
\usepackage{bm}
\usepackage{epsfig}
\usepackage[T1]{fontenc}
\usepackage[latin9]{inputenc}
\usepackage{color}
\usepackage{hyperref}
\usepackage{subfigure}
\usepackage{bm}
\usepackage{ulem}

\def\bea{\begin{eqnarray}}
\def\eea{\end{eqnarray}}

\begin{document}

\title{Evidence for spin-triplet odd-parity superconductivity close to type-II van Hove singularities}
\author{Zi Yang Meng}
\email{zymeng@iphy.ac.cn}
\affiliation{Department of Physics, University of Toronto, Toronto, Ontario M5S 1A7, Canada}
\affiliation{Beijing National Laboratory for Condensed Matter Physics, and Institute of Physics, Chinese Academy of Sciences, Beijing 100190, China}
\author{Fan Yang}
\affiliation{School of Physics, Beijing Institute of Technology, Beijing, 100081, China}
\author{Kuang-Shing Chen}
\affiliation{Institut f\"ur Theoretische Physik und Astrophysik, Universit\"at W\"urzburg, Am Hubland, D-97074 W\"urzburg}
\author{Hong Yao}
\affiliation{Institute for Advanced Study, Tsinghua University, Beijing, 100084, China}
\author{Hae-Young Kee}
\affiliation{Department of Physics, University of Toronto, Toronto, Ontario M5S 1A7, Canada}
\affiliation{Canadian Institute for Advanced Research/Quantum Materials Program, Toronto, Ontario MSG 1Z8, Canada}
\date{\today}

\begin{abstract}
Searching for unconventional Cooper pairing states has been at the heart of superconductivity research since the discovery of BCS superconductors. In particular, spin-triplet odd-parity pairing states were recently revisited due to the possibility of tuning towards topological superconductors. In this context, it is interesting to note a recent proposal that such a spin-triplet pairing instability occurs when the band filling is near van Hove singularities (vHS) associated with momenta away from time-reversal invariant momenta named type-II vHS. However, this result was obtained within a weak coupling renormalization group with Fermi surface patch approximation. To explore superconducting instabilities beyond this weak coupling Fermi surface patch approximation, we perform systematic study on Hubbard model in a two-dimensional square lattice using three different methods: random phase approximation, large-scale dynamical mean field theory simulations with continuous time quantum Monte Carlo (CTQMC) impurity solver, and large-scale dynamical cluster simulations with CTQMC cluster solver. We find, in a wide doping range centered around the type-II van Hove filling,  a two-fold degenerate, spin-triplet, odd-parity $p$-wave pairing state emerges due to repulsive interaction, when the Fermi surface is not sufficiently nested. Possible relevance of our findings to the recently discovered superconductors LaO$_{1-x}$F$_{x}$BiS$_{2}$, Ir$_{1-x}$Pt$_{x}$Te$_{2}$ and proposed doped BC$_{3}$ are also discussed.
\end{abstract}


\maketitle

\section{Introduction}
Identifying superconducting pairing symmetries is at the first step to understand microscopic origins of superconductivity.
While electron-phonon interaction likely leads to a conventional BCS superconductor with $s$-wave spin-singlet pairing symmetry, electronic interactions are well accepted to be responsible for unconventional pairing states including $d$-wave spin-singlet found for example in high Tc curpates~\cite{ZXShen93,Tsuei94,Li1999}. Among the unconventional pairing states, a spin-triplet odd-parity state in electronic systems, an analog of the A-phase in ${}^{3}$He~\cite{vollhardt1990superfluid}, is intriguing. In particular, a chiral spin-triplet with $p+ip$ (short form for $p_{x}\pm ip_{y}$ throughout this paper) odd-parity pairing suggested~\cite{Rice95,Sigrist1999134,Agterberg97,Miyake1999,Nomura2000,Nomura2002a,Nomura2002b,Arita2004} for superconductivity observed in Sr$_2$RuO$_4$~\cite{Maeno94} has been recently revisited~\cite{Raghu2010Sr2RuO4,Kallin2012}, because it can be tuned to a topological superconductor~\cite{schnyder2008,kitaev2009,roy2008,qi2009time,Kashiwaya2011,Ueno2013,Yada2014}, although the experimental and theoretical understanding of the superconducting states in Sr$_{2}$RuO$_{4}$ itself is still under intense debete~\cite{Machida2008,Suderow2009,Zhang2014}, and further work to determine the symmetry of the order parameter is highly desired.
This chiral superconductor also carries Majorana zero mode in vortex cores~\cite{volovik99} which follow non-Abelian statistics~\cite{Ivanov2001}, and could be utilized  for topological quantum computation~\cite{Kitaev03,Nayak08}.
Thus, searching for materials featuring an intrinsic topological superconductor as a new state of matter has been of great interest and considerable efforts have been devoted along this direction~\cite{fu2010odd, qi2010, Raghu2010, raghu2011superconductivity, xiang2012dhlee, Yao13, Chen14}.

In this context, a recent proposal by Yao and Yang~\cite{Yao13} is worthwhile to note.
It was suggested  that an odd-parity $p+ip$ superconductor may appear in two dimensional (2D) electron system when the band filling is close to a particular type of vHS dubbed type-II vHS. Here the type-II vHS means the van Hove (vH) saddle points are located at momenta away from the time-reversal invariant momenta(TRIM), i.e., $\mathbf{K}\neq-\mathbf{K}$ modulo reciprocal lattice vectors. Otherwise, the vHS belongs to type-I.

For systems at vHS and in the limit of weak repulsive interactions, singlet and triplet pairings generically compete on equal footing as both of their superconducting susceptibilities have square of logarithmical divergence. But there exist qualitative distinctions between type-I and type-II vHS~\cite{Yao13}. At type-II vH saddle points, singlet and triplet pairings are both allowed, in contrast with strong suppression of triplet pairing around type-I vHS by the Pauli exclusion principle, as the associated momenta are TRIM~\cite{dzyaloshinskii1987maximal,schulz1987superconductivity,lederer1987antiferromagnetism, furukawa1998truncation,nandkishore2012chiral,kiesel2012competing,wang2012functional}. RG analysis~\cite{Yao13} showed  in the limit of weak repulsive interactions, triplet pairing is most favored when the Fermi surface is not sufficiently nested. For type-II vHS system that respects tetragonal symmetry~\cite{Yao13} ({\it i.e.} square lattice) or hexagonal symmetry~\cite{Chen14} ({\it i.e.} honeycomb lattice), topological superconductivity (either chiral $p+ip$ pairing or time-reversal invariant $Z_2$) could occur.

However, the RG analysis of superconductivity instability is only perturbative and reliable in the limit of weak interaction strength. Furthermore, because the Fermi surface (FS) was approximated as several patches around vHS momenta,  the FS curvature effects were ignored.
Whether spin-triplet $p+ip$ pairing will survive at intermediate or strong electronic interactions beyond FS patch approximation is an open and urgent question to answer.
In this paper, we directly address this question in 2D type-II vHS system with advanced numerical methods, such that the correlation effects can be captured more accurately. Specifically, we employ three different methods which treat electronic correlations at different level. 
Since the ferromagnetic fluctuation is essential for the spin-triplet pairing in ${}^{3}$He A-phase~\cite{Anderson73}, one could speculate that systems with strong ferromagnetic instability could be a good candidate for spin-triplet superconductivity, where these two instabilities compete.
Thus we study the competition among  spin-singlet and spin-triplet pairing states in addition to ferromagnetic state (antiferromagnetic instability is also checked and found to be small).
Results from these three methods converge to a coherent picture that the spin-triplet with $p+ip$ pairing superconductor is indeed stabilized by repulsive interaction in 2D system with type-II vHS, due to the enhanced ferromagnetic fluctuations, from weak to strong couplings.

The methods we employed are random phase approximation (RPA)~\cite{bickers1989conserving,RPA1,RPA2,RPA3,RPA4,RPA5,RPA6}, large-scale dynamical mean field theory (DMFT) incorporated with parquet simulations~\cite{Meng14, iQIST2014}, and large-scale dynamical cluster approximation (DCA) simulations~\cite{Maier05,Chen12,Chen13b}. In order to go beyond the local nature of DMFT in studying the pairing symmetries, we employed the recently developed DMFT and parquet (DMFT+Parquet) formalism~\cite{Meng14,iQIST2014}, in which we make use of the local vertex and correlation functions measured in DMFT/CTQMC simulations, and introduces the momentum-dependence into these two-particle quantities via two-particle self-consistent equations, i.e., Bethe-Salpeter and parquet equations~\cite{Dominicis64a, *Dominicis64b}.

In the large-scale DCA simulations, a 16-site momentum-space cluster~\cite{Maier05,Chen12,Chen13b} is used. The cluster is solved exactly via CTQMC~\cite{Rubtsov05} and coupled to the mean-field bath self-consistently via iterations~\cite{Hettler98,Hettler00}. We have designed the tight-binding parameters such that the vH momenta are among the cluster momentum points. By measuring the two-particle vertex and correlation functions directly on the cluster, we capture the interaction effects close to vHS.

In this paper, the DMFT+Parquet formalism is for the first time being applied to one-band model. Furthermore, the consistent results obtained from DMFT+Parquet and DCA simulations provide the first benchmark to DMFT+Parquet to confirm its credibility for future usage in addressing the interacting instabilities in other strongly correlated electron systems.

The paper is organized as follows. Sect. II outlines the model and previous results based on the RG analysis. Sect. III explains the three methods employed here and discusses their corresponding results. In the DMFT+Parquet part, we explain how the momentum dependence is introduced into two-particle vertex and correlation functions. In the DCA part, we demonstrate that 16-site cluster is capable of capturing the vH momenta. We end with discussion about relevances of the present study to experiments and suggestion for future direction in Sect. IV.

\section{Model and RG analysis}
\begin{figure}[t]
\centering{}
\includegraphics[width=3.8in]{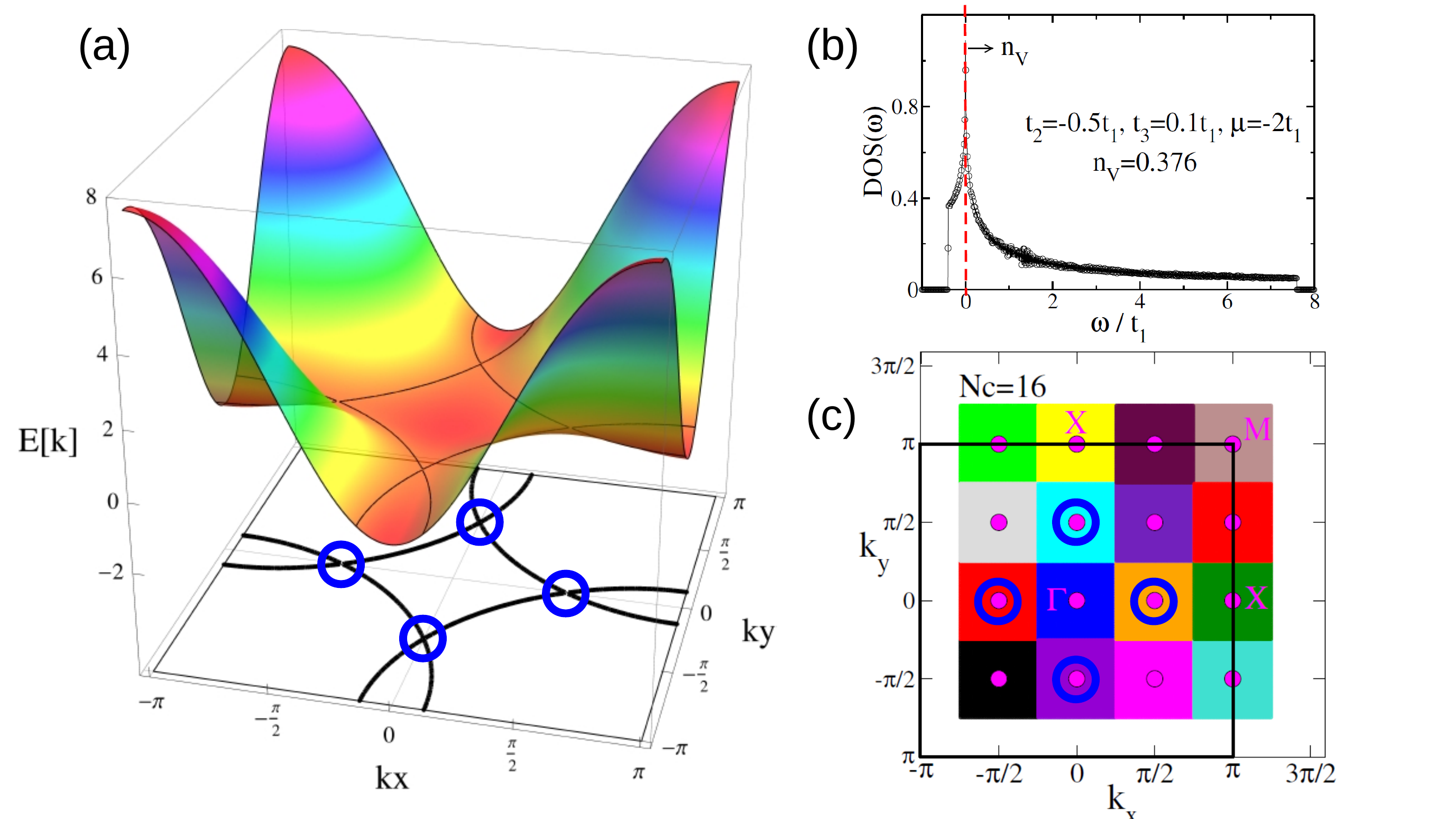}
\caption{(color online) (a) Energy dispersion of the non-interacting system, with parameters $t_{2}=-0.5t_{1}$, $t_{3}=0.1t_{2}$, $\mu=-2t_{1}$. Fermi surface is highlighted by the black contour line and its projection to BZ at the bottom. Blue circles represent the position of vH momenta at $\mathbf{K}=(\pm\pi/2,0)$ and $(0,\pm\pi/2)$. (b) Density of the states, $\text{DOS}(\omega)$, the VHS is located at the Fermi level and the corresponding band filling $n_{V}=0.376$, as indicated by red dashed line. (c) The BZ patches of $N_{c}=16$ cluster for DCA simulations. The self-energy in each patch is approximated by the self-energy at the cluster momentum point (solid pink dot). The vH momenta (blue circles) are captured by $N_{c}=16$ cluster.}
\label{fig:dispersion}
\end{figure}

We consider the Hubbard model on 2D square lattice,
\begin{equation}
H=-\sum_{i,j,\sigma}t_{i,j}c^{\dagger}_{i,\sigma}c_{j,\sigma} -\sum_{i,\sigma}\mu c^{\dagger}_{i,\sigma}c_{i,\sigma} + \sum_{i}U n_{i,\uparrow}n_{i,\downarrow},
\label{eq:hamiltonian}
\end{equation}
where $c^{\dagger}_{i,\sigma}$ ($c_{i,\sigma}$) are electron creation (annihilation) operators with spin $\sigma=\uparrow,\downarrow$ at site $i$. $t_{i,j}=t_{1}, t_{2}, t_{3}$ are the hoppings between first, second, and third neighboring sites, respectively. $U$ is the onsite repulsive interaction. For the non-interacting system, when $|t_{1}+2t_{2}|>4t_{3}$ and $\mu=4(t_{2}-t_{3})$, the FS possesses type-I vH saddle points at the boundary of Brillouin zone (BZ), $\mathbf{K}=(0,\pi)$ and $(\pi,0)$, these momenta are TRIM. 
Here, we focus on the type-II vHS~\cite{Yao13,Chen14}, in which $\mathbf{K}\neq-\mathbf{K}$ is satisfied. In the 2D Hubbard model we considered, the type-II vHS can be realized for $|t_{1}+2t_{2}|<4t_{3}$ and $\mu=(t_{1}+2t_{2})^{2}/(4t_{3})-2t_{1}$\cite{Yao13}. For the parameter set $t_{2}=-0.5t_{1}$, $t_{3}=0.1t_{1}$, chemical potential $\mu=-2t_{1}$ and the four type-II vH saddle points are located at $\mathbf{K}=(\pm\pi/2,0)$ and $(0,\pm\pi/2)$, as shown by the blue circles in the Fig.~\ref{fig:dispersion} (a).

For a generic FS without perfect nesting and in the limit of weak interactions, Cooper pairing is always the leading instability of the FS since pairing susceptibility of non-interacting fermions is more logarithmically divergent than any other particle-hole channel in low energy or low temperature~\cite{schrieffer1999theory}. To determine which pairing channel is most favored by weak repulsive interactions, perturbative RG analysis was often employed~\cite{dzyaloshinskii1987maximal,schulz1987superconductivity,lederer1987antiferromagnetism, furukawa1998truncation,Raghu2010,raghu2011superconductivity}, for a review, see Ref.~\onlinecite{Metzner2012}. For 2D systems at vHS, the DOS is logarithmically divergent, which is mainly due to the electrons around vH saddle points. In the limit of weak interactions, fermions around the FS dominate the low energy physics. Consequently, it is reasonable to neglect electrons far away from vH saddle points and only focus on electrons in patches around vH momenta~\cite{dzyaloshinskii1987maximal,schulz1987superconductivity,lederer1987antiferromagnetism, furukawa1998truncation,nandkishore2012chiral}. Within the patch approximation, there are six independent interactions labeled by $g_i$, $i=1,\cdots,6$. RG flow of these interactions was analyzed in Ref.\onlinecite{Yao13} and it was found that triplet $p$-wave pairing is more favored than pairings in all singlet channels when the FS is not close to perfect nesting. Heuristically, the favoring of triplet pairing in the limit of weak interactions is due to the dominance of ferromagnetic fluctuations of non-interacting fermions when the FS is not close to perfect nesting. Because the system in question respects the tetragonal symmetry, two $p$-wave pairings are degenerate and consequently topological $p+ip$ pairing is expected based on general arguments in analyzing the Ginzburg-Landau free energy~\cite{Sigrist91,cheng2010stable}.

Given the RG analysis is only valid at weak coupling limit, we employ more advanced techniques in a systematic manner, with the level of exactness progressively increasing, to study correlation effects on the superconductivity instabilities from weak to strong interactions. Below we present the three different numerical techniques and their corresponding results.


\section{Numerical methods and Results}
\subsection{Random Phase Approximation}
\label{sec:RPA}
We first perform a RPA based study~\cite{bickers1989conserving,RPA1,RPA2,RPA3,RPA4,RPA5,RPA6}.  In this approach, the particle-hole charge or magnetic susceptibilities are calculated in the RPA level first, with the vertex functions replaced by the bare interaction $U$. Then, through exchanging the charge or magnetic fluctuations (whose propagators are represented by corresponding susceptibilities), the electrons near the FS acquire effective attractions. Finally, by solving the linearized gap equation near $T_{c}$, one obtains the leading pairing symmetries and their corresponding $T_{c}$.

\begin{figure}
\includegraphics[width=3.3in]{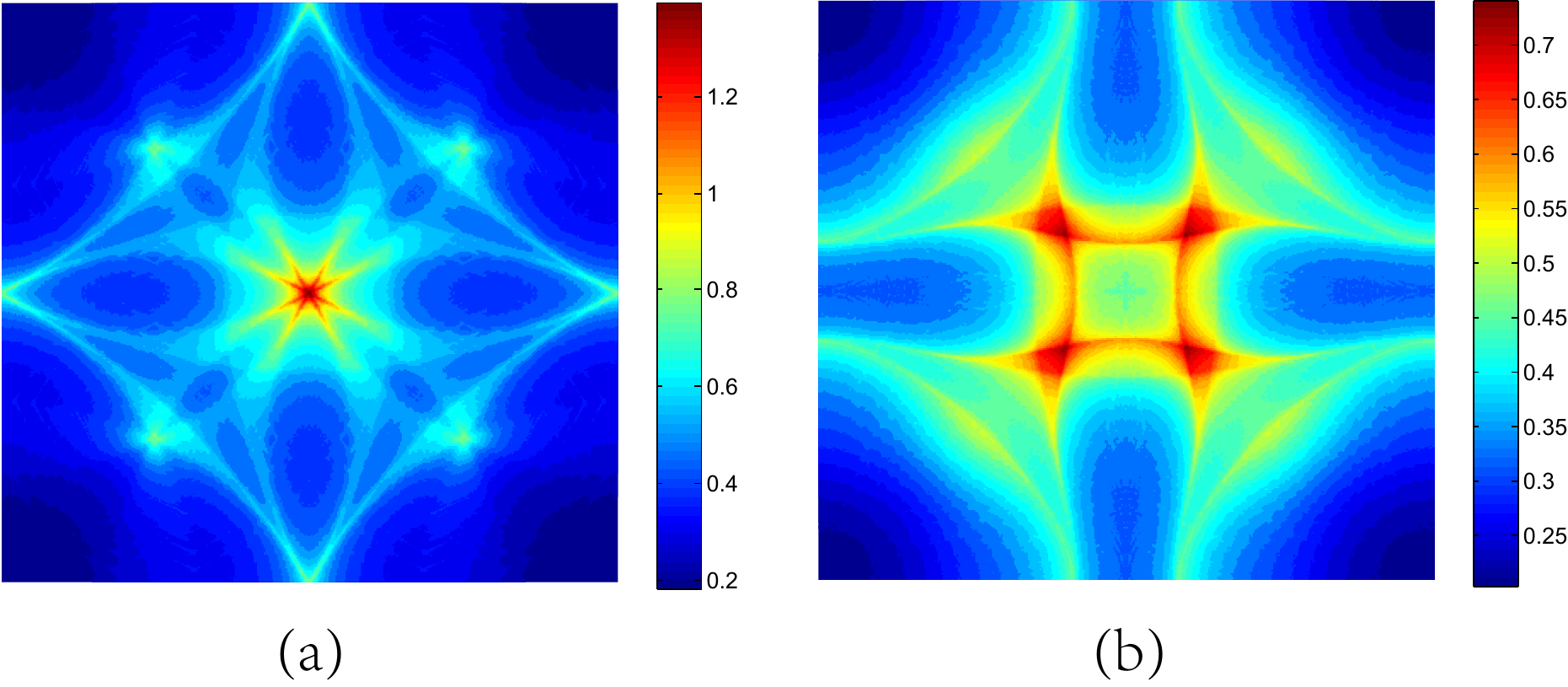}
\caption{(color online) Distribution of the bare susceptibilities $|\chi^{ph}_{0}\left(\mathbf{q}\right)|$ in the BZ for the filling of (a) $n_{V}=0.376$ and (b) $n=0.5$. The $\chi^{ph}_{0}(\mathbf{q})$ is peaked close to $\Gamma=\mathbf{q}=(0,0)$.} \label{fig:sus}
\end{figure}

The bare susceptibility in the particle-hole channel $\chi^{ph}_{0}\left(\mathbf{q},\tau\right)$ (for $U=0$) of the model is given by,
 \begin{eqnarray}
 \chi^{ph}_{0}\left(\mathbf{q},\tau\right)\equiv
 \frac{-1}{N}\sum_{\mathbf{k_{1},k_{2}}}\left<T_{\tau}c^{\dagger}(\mathbf{k_{1}},\tau)
 c(\mathbf{k_{1}+q},\tau)\right.\nonumber\\ \left. c^{\dagger}(\mathbf{k_{2}+q},0)c(\mathbf{k_{2}},0)\right>_0.
 \end{eqnarray}
 Its zero-frequency component is evaluated as,
 \begin{eqnarray}
 \chi^{ph}_{0}\left(\mathbf{q},i\omega=0\right)&\equiv&\chi^{ph}_{0}\left(\mathbf{q}\right)\nonumber\\&=&\frac{1}{N}\sum_{\mathbf{k}}
  \frac{n_{F}(\varepsilon_{\mathbf{k}})-n_{F}(\varepsilon_{\mathbf{k+q}})}
 {\varepsilon_{\mathbf{k}}-\varepsilon_{\mathbf{k+q}}}.
 \end{eqnarray}
The distribution of $|\chi^{ph}_{0}\left(\mathbf{q}\right)|=-\chi^{ph}_{0}\left(\mathbf{q}\right)$ in the BZ for the band filling close to the vH filling denoted by $n_V$ is shown in Fig.~\ref{fig:sus} (a) and for the filling $n=0.5$ in Fig.~\ref{fig:sus} (b) for comparison.
Using the above tight binding parameters, $n_V=0.376$. Fig.~\ref{fig:sus} (a) shows that at the vH filling, the susceptibility peaks at the $\Gamma$-point, which suggests dominant ferromagnetic spin or charge fluctuations in the system. When the filling gradually increases, the momenta with maximum susceptibility deviate from the $\Gamma$-point slowly. As shown in Fig.~\ref{fig:sus} (b), the bare susceptibility for the filling $n=0.5$ peaks at small momenta around the $\Gamma$-point, again suggesting dominant ferromagnetic-like fluctuations.

When the Hubbard interaction is switched on, the particle-hole charge and magnetic susceptibilities are given (in the RPA level) by,
\begin{equation}
\chi^{c/m}_{ph}\left(\mathbf{q}\right)=
\left[1\mp U\chi^{ph}_{0}\left(\mathbf{q}\right)\right]^{-1}\chi^{ph}_{0}
\left(\mathbf{q}\right),\label{RPA}
\end{equation}
where the ``$-$"(``$+$") represents for charge (magnetic) channels. It's clear that the repulsive Hubbard-interaction suppresses $\chi^{(c)}$ while it enhances $\chi^{(m)}$. Thus the magnetic fluctuations dominate in the system.

\begin{figure}
\includegraphics[width=2.4in]{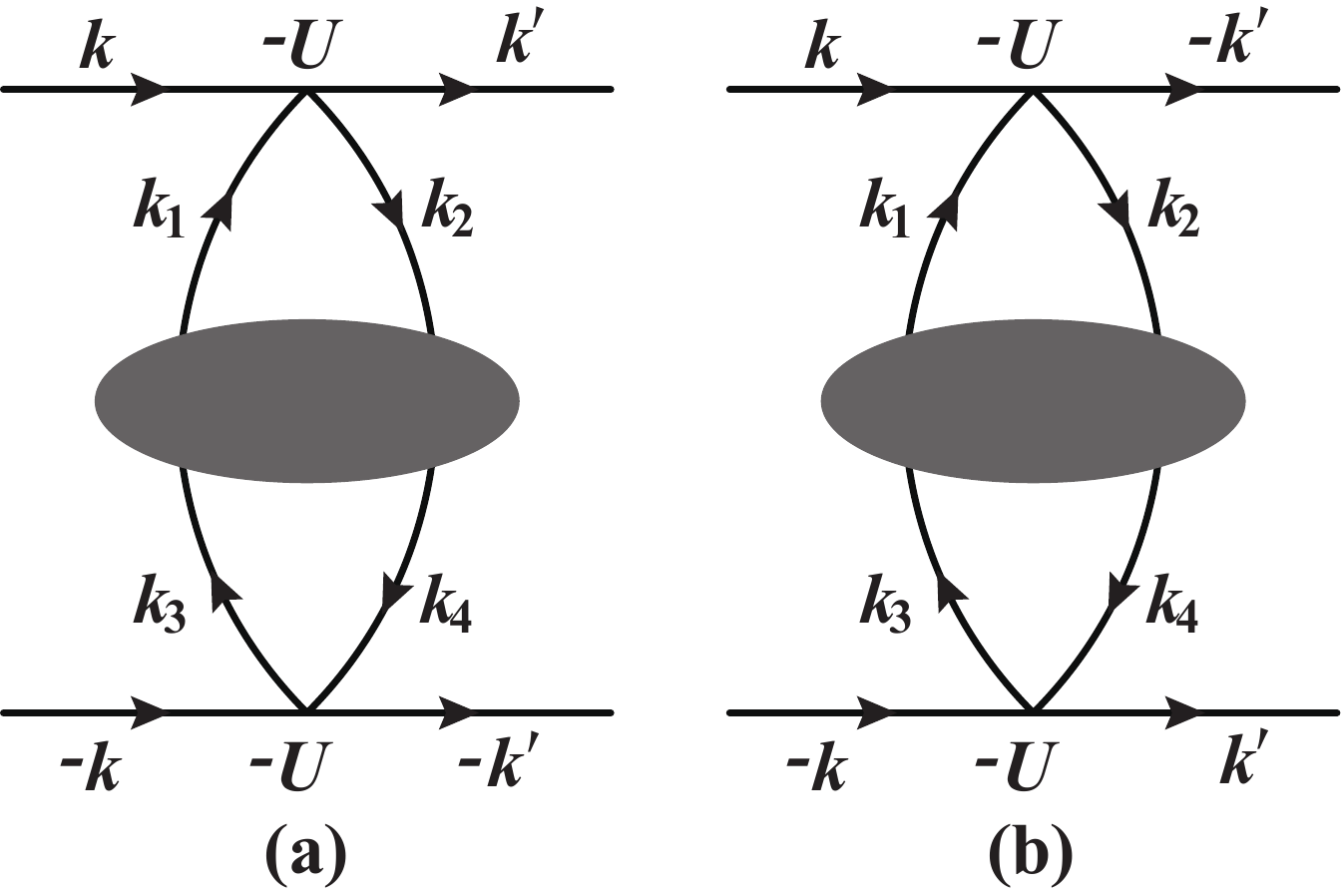}
\caption{RPA diagrams of the two second order processes which contribute the effective interaction Eq.~\ref{effective_interaction}, wherein a cooper pair with momenta $(\mathbf{k,-k})$ are scattered to the momenta $(\mathbf{k',-k'})$ by the magnetic fluctuation represented by the renormalized magnetic susceptibility (shaded ellipses).}\label{fig:rpa}
\end{figure}

Through exchanging ferromagnetic fluctuations between each other, the electrons near the FS acquire an effective attraction, which leads to cooper pairing in the system. Such processes are illustrated by the Feynman diagrams shown in Fig.~\ref{fig:rpa} (a) and (b), from which one obtains the following effective interaction
\begin{equation}
 V_{eff}=\frac{1}{N}\sum_{\mathbf{kk'}}V(\mathbf{k,k'})c^{\dagger}
 (\mathbf{k})c^{\dagger}(-\mathbf{k})c(-\mathbf{k}')c(\mathbf{k}').\label{effective_interaction}
 \end{equation}

Here the effective vertex $V(\mathbf{k,k'})$ takes the form of
 \begin{eqnarray}
V^{s}_{pp}(\mathbf{k,k'})&=&U+
\frac{U^2}{4}\left[\chi^{c}_{ph}\left(\mathbf{k-k'}\right)-3\chi^{m}_{ph}\left(\mathbf{k-k'}\right)\right]\nonumber\\
&&+\frac{U^2}{4}\left[\chi^{c}_{ph}\left(\mathbf{k+k'}\right)-3\chi^{m}_{ph}\left(\mathbf{k+k'}\right)\right]
\label{eq:RPA_singlet},
 \end{eqnarray}
for particle-particle singlet pairing, while in the particle-particle triplet channel, it is given by
\begin{eqnarray}
V^{t}_{pp}(\mathbf{k,k'})&=&\frac{U^2}{4}\left[\chi^{c}_{ph}\left(\mathbf{k-k'}\right)+\chi^{m}_{ph}\left(\mathbf{k-k'}\right)\right]\nonumber\\&&-
\frac{U^2}{4}\left[\chi^{c}_{ph}\left(\mathbf{k+k'}\right)+\chi^{m}_{ph}\left(\mathbf{k+k'}\right)\right].
 \end{eqnarray}
Solving the following linearized gap equation
  \begin{equation}
 -\frac{1}{(2\pi)^2}\oint_{FS}
dk'_{\Vert}\frac{V(\mathbf{k,k'})}{v_{F}(\mathbf{k'})}\Delta(\mathbf{k'})=\lambda
 \Delta(\mathbf{k}),
 \label{eq:eigenvalue_Tc}
\end{equation}
one can obtain the leading pairing symmetry in the system. Here $v_{F}(\mathbf{k'})$
is the Fermi velocity and $k'_{\Vert}$ represents the component along the FS. The leading pairing symmetry of the system is thus determined by the leading eigenvalue (LEV) $\lambda$ of Eq.~\ref{eq:eigenvalue_Tc}. The superconducting critical temperature $T_c$ is determined via $T_{c}\approx t_{1} e^{-1/\lambda}$, where $\lambda$ is the leading eigenvalue,  and the corresponding normalized eigenvector $\Delta(\mathbf{k})$ represents the relative gap function on the FS.

\begin{figure}
\includegraphics[width=3.3in]{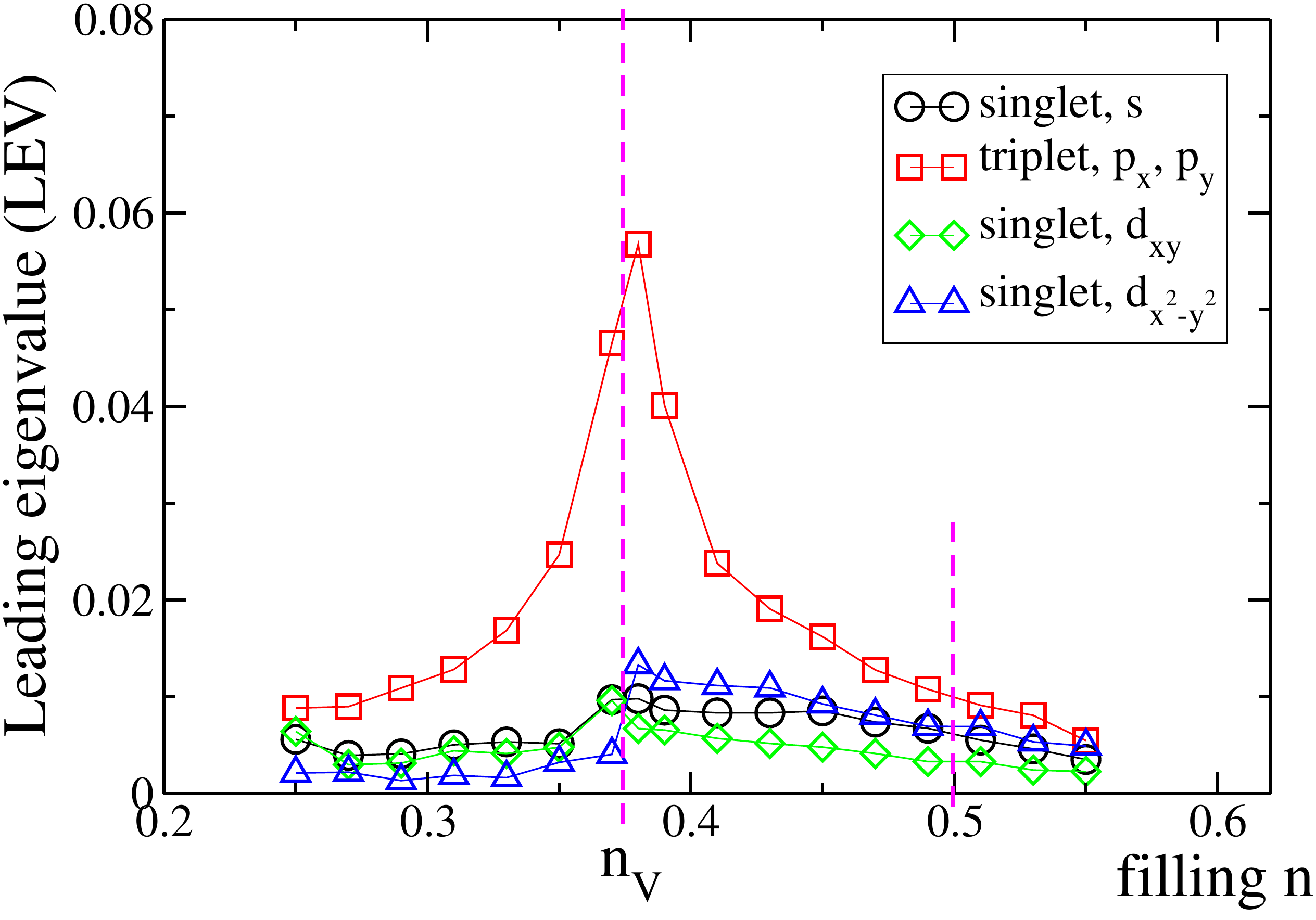}
\caption{(color online) The doping dependence of the LEV $\lambda$ of Eq.~\ref{eq:eigenvalue_Tc} for the four possible pairing symmetries in the square lattice, i.e. the s-wave, the d$_{xy}$, the d$_{x^2-y^2}$ and the degenerate $p$-wave (i.e. $p_x$ and $p_y$). Here, a weak interaction parameter $U=0.5t_1$ and a low finite temperature $k_BT=10^{-4}t_1$ are adopted to avoid the divergence of the magnetic susceptibility near the vHS. The pink dashed lines point out the fillings at which the leading eigenvectors are presented in Fig.~\ref{fig:gap}.}\label{fig:egn}
\end{figure}

\begin{figure}
\includegraphics[width=3in]{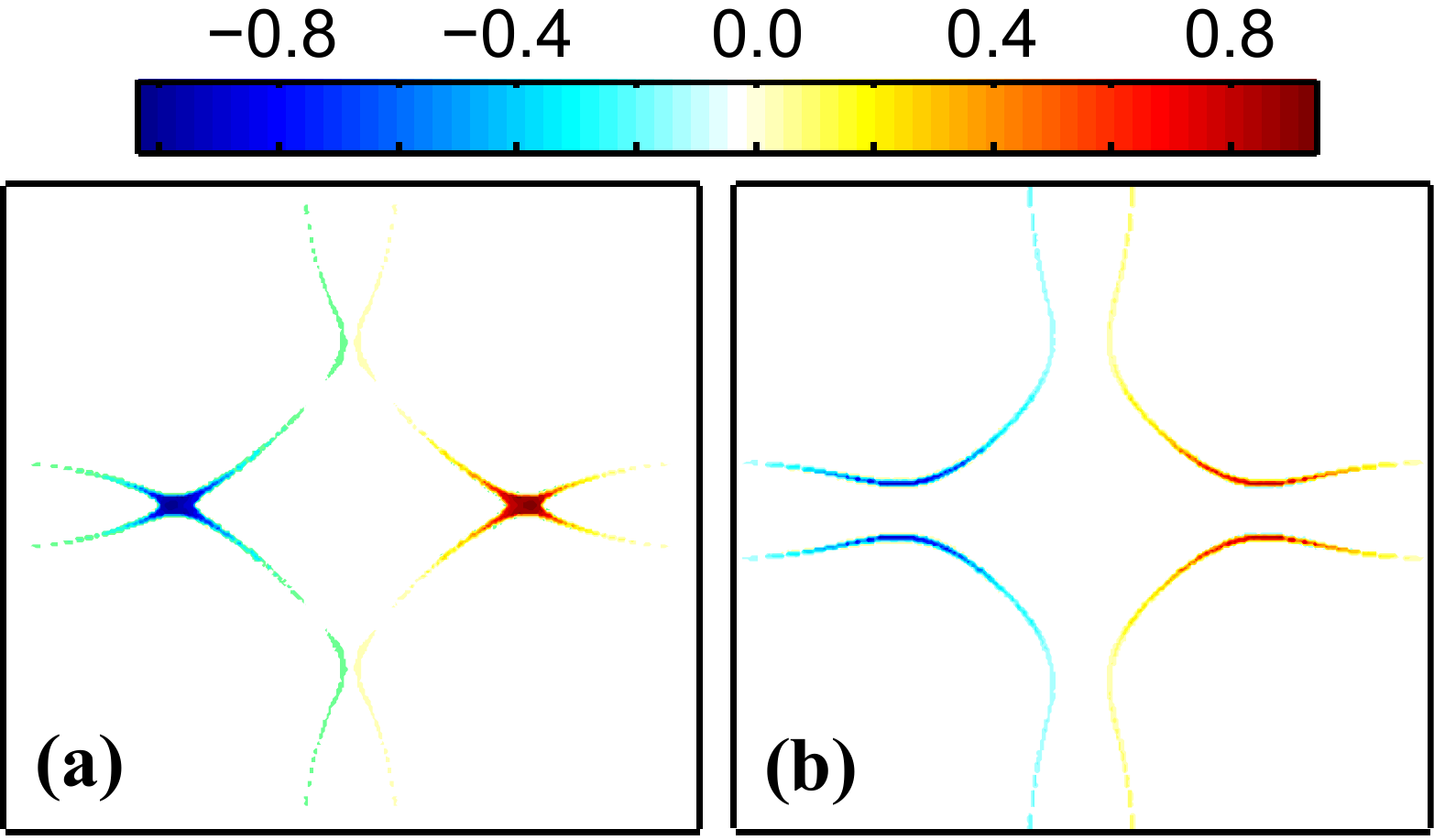}
\caption{(color online) The leading pairing gap function $\Delta(\mathbf{k})$ of $p_x$ symmetry for the vH filling $n_V=0.376$ ((a)) and filling $n=0.5$ ((b)). The other one of $p_y$ symmetry can be obtained from the present one by a 90$^o$ rotation.}\label{fig:gap}
\end{figure}

The doping-dependence of the leading eigenvalue $\lambda$ for all the four possible pairing symmetries in the square lattice, {\it i.e.} the $s$-wave, the $d_{xy}$, the $d_{x^2-y^2}$, and the degenerate $p$-wave (i.e. $p_x$ and $p_y$), is shown in Fig.~\ref{fig:egn}. Here, we have adopted a relatively weak interaction parameter $U=0.5t_1$ and a low but finite temperature $T=10^{-4}t_1$ to avoid the divergence of the magnetic susceptibility near the vHS. From Fig.~\ref{fig:egn}, in the whole doping regime $n\in(0.25,0.55)$ near the vH filling $n_V=0.376$, the leading pairing symmetry of the system is always the degenerate $p$-wave, which possesses the highest $T_c$ near $n_V$. As for the sub-leading pairing symmetry, it will shift from the $s$-wave below the vH filling $n_V$ to the $d_{x^2-y^2}$-wave above $n_V$. One of the leading pairing gap function, which is of $p_x$ symmetry, is shown in Fig.~\ref{fig:gap} (a) for filling $n_V=0.376$ and in Fig.~\ref{fig:gap} (b) for filling $n=0.5$. The other one of $p_y$ symmetry can be obtained from the shown one by a 90$^\circ$ rotation. The two $p$-wave pairing states have the same $T_c$. When the temperature is lowered below $T_c$, energy minimization for the effective Hamiltonian including Eq.~\ref{effective_interaction} as the interaction part leads to a $p+ip$ pairing.

Note that the RPA approach fails at the VH doping level because the divergent DOS, and hence the divergent spin susceptibility there urges the formation of long-range FM magnetic order even for vanishingly weak U, which closes the door to the formation of SC. Therefore, the RPA only applies to doping levels away from the VHS.

\subsection{DMFT+Parquet}
By construction, RPA works better for weak interactions. For intermediate and strong interactions, better treatments of correlations are desired since the two-particle irreducible vertex functions in particle-particle (pp) $\Gamma^{s/t}_{pp}$ and particle-hole (ph) $\Gamma^{c/m}_{ph}$ channels ($s/t$ refers to singlet and triplet and $c/m$ refers to charge and magnetic channels) acquire non-trivial momentum and frequency dependence, whereas in RPA they are approximated by bare Hubbard $U$. In this section, we employ the recently developed DMFT+Parquet formalism~\cite{Meng14,iQIST2014} to explore the superconductivity instabilities of the system with stronger interactions. In this approach, the frequency dependence of the the vertex and correlation functions $\chi^{s/t}_{pp}$ and $\chi^{c/m}_{ph}$ are captured exactly on the impurity site, and the momentum dependence is introduced by two-particle diagrammatical techniques.

The interaction Hamiltonian Eq.~\ref{eq:hamiltonian} is solved by DMFT with hybridization expansion CTQMC impurity solver~\cite{Werner06,Haule07,Huang12a}. This scheme maps the original correlated lattice problem into a quantum impurity embedded into a self-consistently determined bath. The self-consistency is achieved while the single particle Green's functions in imaginary time, $G(\tau)$, or in matsubara frequency, $G(\omega)$ are identical both on the impurity and in the bath. The interaction strength is chosen as $U=2t_{1}$ and $4t_{1}$, and the band filling is $n=0.4$, close to the $n_{V}$. We can achieve temperatures as low as $T=0.02t_{1}$ before serious minus-sign problem occurs.

\begin{figure}[t]
\centerline{
\includegraphics[width=3.4in]{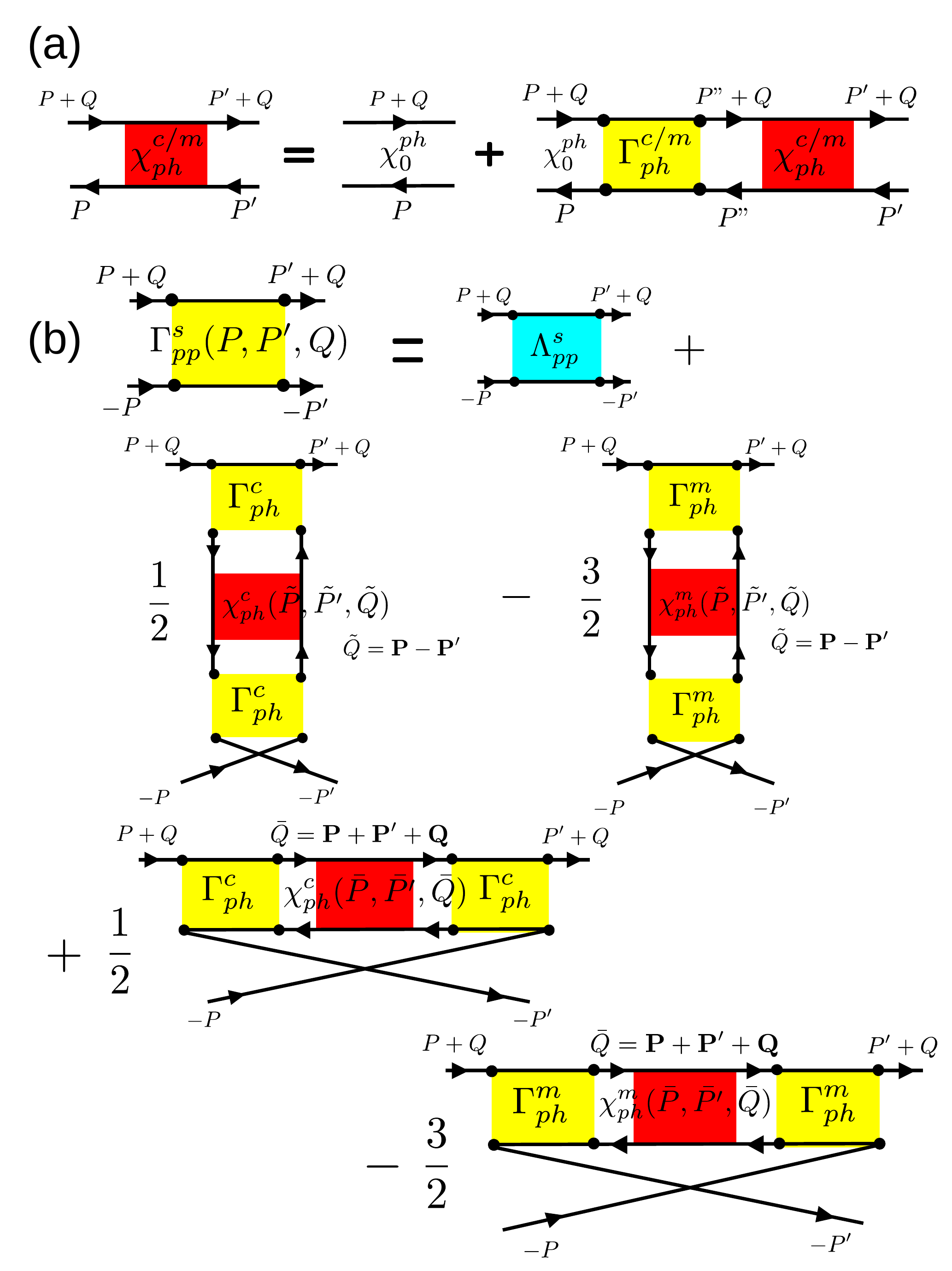}}
\caption{(color online) (a) Bethe-Salpeter equations in the particle-hole charge/magnetic channels. $\chi^{c/m}_{ph}(P,P',Q)$ are the two-particle correlation functions and $\Gamma^{c/m}_{ph}(P',P",Q)$ are the irreducible vertex functions, $\chi^{ph}_{0}(P,Q)$ is the bare bubble. (b) Parquet equation of the irreducible particle-particle singlet vertex function, $\Gamma^{s}_{pp}(P,P',Q)$. It is decomposed into fully irreducible vertex $\Lambda^{s}_{pp}$ and cross channel contributions from particle-hole charge/magnetic vertex ladders $\Phi^{c/m}_{ph}$. The vertex ladders are products between irreducible vertex functions, $\Gamma$, and two-particle correlation function, $\chi$, in the same channel, with the internal momentum-frequency indices integrated out, but the momentum-frequency transfer $\tilde{Q}$ and $\bar{Q}$ determined by the indices of the irreducible vertex functions in the LHS of the equation.}
\label{fig:parquet_pp}
\end{figure}

To obtain momentum-dependence in the local two-particle vertex and correlation functions measured on the impurity site, we make use of Bethe-Salpeter and parquet equations (schematically shown in Fig.~\ref{fig:parquet_pp} (a), (b)). Parquet equations~\cite{Dominicis64a, *Dominicis64b} relate the irreducible vertex function in one interaction channel with the irreducible vertex functions in other channels~\cite{Bickers91,Yang09,Rohringer12,Tam13,Chen13a,Chen13b}. There are four channels here: the particle-hole charge (ph-c), particle-hole magnetic (ph-m), particle-particle singlet (pp-s) and particle-particle triplet (pp-t). The two-particle diagrammatical calculation scheme has been explained in detail in Refs.~\onlinecite{Meng14,iQIST2014}, here we outline the main procedure.


The DMFT/CTQMC simulations provide the lattice single-particle Green's function $G(P)$ and the local, ph-c/m two-particle correlation functions $\chi^{c/m}_{ph}(\omega,\omega',\nu)$. The ph bubble term $\chi^{ph}_{0}(P,Q)$ (bare susceptibility in Sec.~\ref{sec:RPA}) can be constructed as $\chi^{ph}_{0}(P,Q)=-N\beta G(P)G(P+Q)$ with $N$ the lattice size and $P\equiv(\mathbf{k},\omega)$, $Q\equiv(\mathbf{q},\nu)$.

We extract the local irreducible vertex functions, $\Gamma^{c/m}_{ph}(\omega,\omega'',\nu)$ from the Bethe-Salpeter equation (Fig.~\ref{fig:parquet_pp} (a)) on the impurity, $\chi^{c/m}_{ph}(\omega,\omega',\nu)=\chi^{ph}_{0}(\omega,\nu)+\chi^{ph}_{0}(\omega,\nu)\sum_{\omega''}\Gamma^{c/m}_{ph}(\omega,\omega'',\nu)\chi^{c/m}_{ph}(\omega'',\omega',\nu)$, then use the bubble term $\chi^{ph}_{0}(P,Q)$ to insert the momentum transfer $\mathbf{q}$ into the two-particle correlation function, $\chi^{c/m}_{ph}(\omega,\omega',Q=(\mathbf{q},\nu)) = \big[[\frac{1}{N\beta}\sum_{P}\chi^{ph}_{0}(P,Q)]^{-1} - \Gamma^{c/m}_{ph}(\omega,\omega',\nu)\big]^{-1}$.

Next we make use of the parquet equations, for the pp-s channel, it is
shown in Fig.~\ref{fig:parquet_pp} (b) (the equations in pp-t and ph-c/m channels are given in Ref.~\onlinecite{Meng14}): $\Gamma^{s}_{pp} (P,P',Q)$ is decomposed into 
cross channel contributions via the ph vertex ladders $\Phi^{c/m}_{ph} = \Gamma^{d/m}_{ph} \star \chi^{c/m}_{ph} \star \Gamma^{c/m}_{ph}$, where $\star$ represents the convolution both in momentum and frequency. We use $\Gamma^{c/m}_{ph} (\omega,\omega',\nu)$ and $\chi^{c/m}_{ph}(\omega,\omega',{\tilde Q})$ to approximate $\Gamma^{c/m}_{ph} (P,P',Q)$ and $\chi^{c/m}_{ph}(P,P',{\tilde Q})$ in the ph ladders $\Phi^{c/m}_{ph}$. 
As ${\tilde Q} = P-P'$ or $P+P'+Q$, the momentum dependence in $\Gamma^{s}_{pp} (P,P',Q)$ is achieved. Similar procedure is employed to get $\Gamma^{t}_{pp}(P,P',Q)$ and $\Gamma^{c/m}_{ph} (P,P',Q)$. One can then iterate $\Gamma^{s/t}_{pp}(P,P',Q)$ and $\Gamma^{c/m}_{ph}(P,P',Q)$ back to Bethe-Salpeter and parquet equations to successively generate the higher order two-particle quantities. Here we only keep the first order results.

It is interesting to notice the inherent relation between the RPA calculation in previous section and the DMFT+Parquet technique applied here. Parquet equation in Fig.~\ref{fig:parquet_pp} (b) will reduce to Eq.~\ref{eq:RPA_singlet} if one replaces the irreducible vertex functions $\Gamma^{c/m}_{ph}(P,P',Q)$ with the bare interaction $U$. This means the RPA results provide the lowest order results of the interaction effect, and once the interaction effect becomes more and more important, i.e., the vertex functions start to have complicated structure in momentum and frequency, more advanced techniques becomes necessary to provide both qualitatively and quantitatively correct physics.

\begin{figure}[tp!]
\centerline{
\includegraphics[width=3.5in]{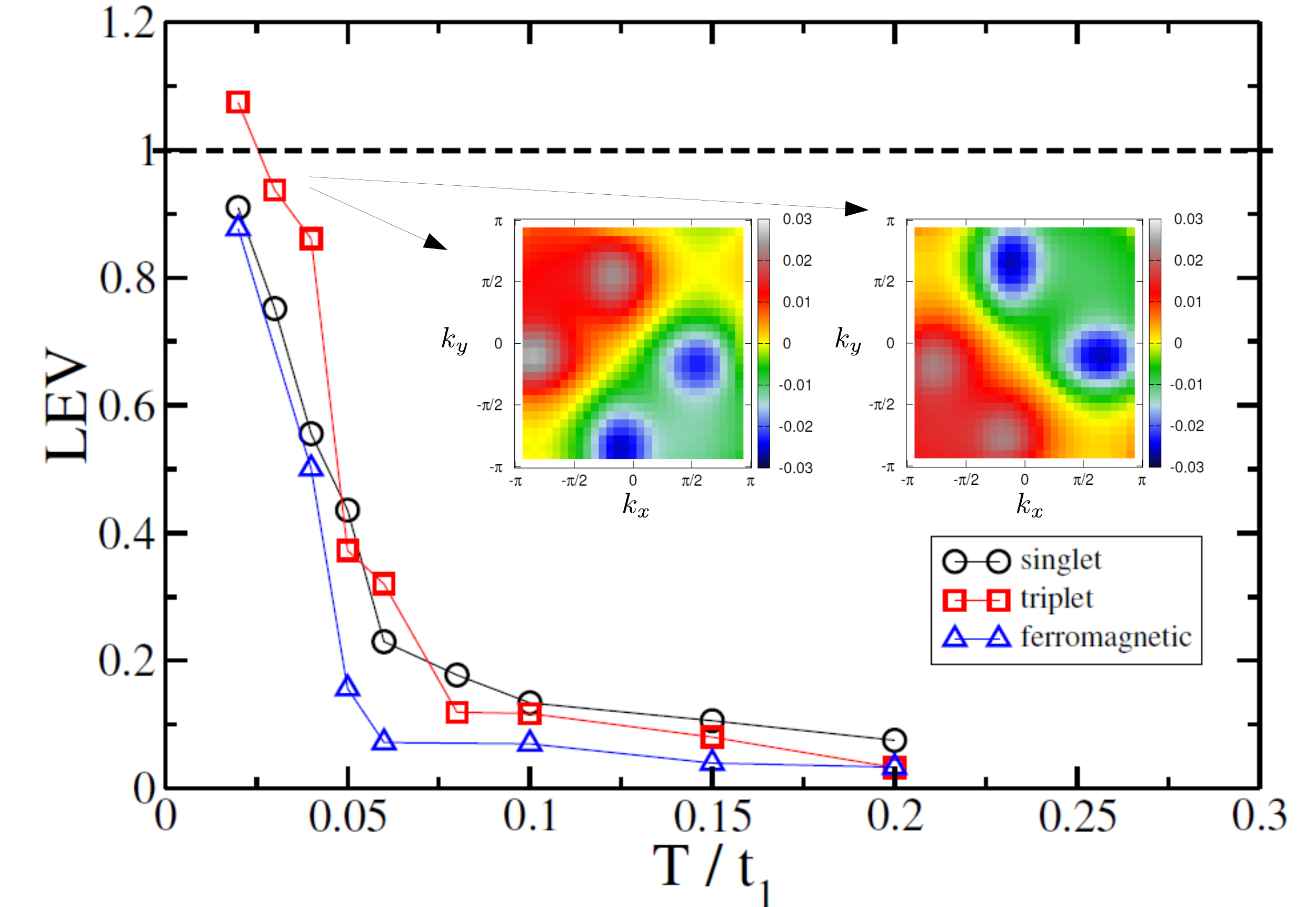}}
  \caption{(color online) LEVs close to the type-II vHS with $t_{2}=-0.5t_{1}$, $t_{3}=0.1t_{1}$, $n=0.4$ and $U=2t_{1}$. As temperature goes down, triplet pairing wins over the singlet pairing, ferromagnetic instability and becomes the leading instability of the system. The corresponding leading eigenvectors are two-fold degenerate, one has $p'_{x}=-p_{x}-p_{y}$ symmetry (left inset) and the other has $p'_{y}=-p_{x}+p_{y}$ symmetry (right inset). The regions with larger amplitude of $\phi(P)$ are close to the vH momenta $\mathbf{K}=(\pm\pi/2,0)$ and $(0,\pm\pi/2)$.}
  \label{fig:single_site_U2}
\end{figure}

With irreducible vertex functions $\Gamma^{s/t}_{pp}(P,P',Q)$ obtained, pairing instabilities of the system can be accessed, one can solve the eigen-equations of the pairing matrix, constructed by the irreducible vertex function and the bare bubble term, in pp-s/t channels,
\begin{equation}
\sum_{P'}\Gamma^{s/t}_{pp}(P,P',Q)\chi^{pp}_0(P',Q)\phi(P') = \lambda \phi(P),
\label{eq:pairing_eigen_equation}
\end{equation}
and analyze the LEV $\lambda$ and the leading eigenvector $\phi(P)$ (note this is the advanced version of Eq.~\ref{eq:eigenvalue_Tc}). As temperature goes towards the transition temperature $T_{c}$, $\lambda \to 1$, and the corresponding $\phi(P)$ reveals the momentum-dependence of the gap function~\cite{schrieffer1999theory,Chen13a}. Similar analysis can be performed in the ph-c/m channels.

\begin{figure}[tp!]
\centerline{
\includegraphics[width=3.5in]{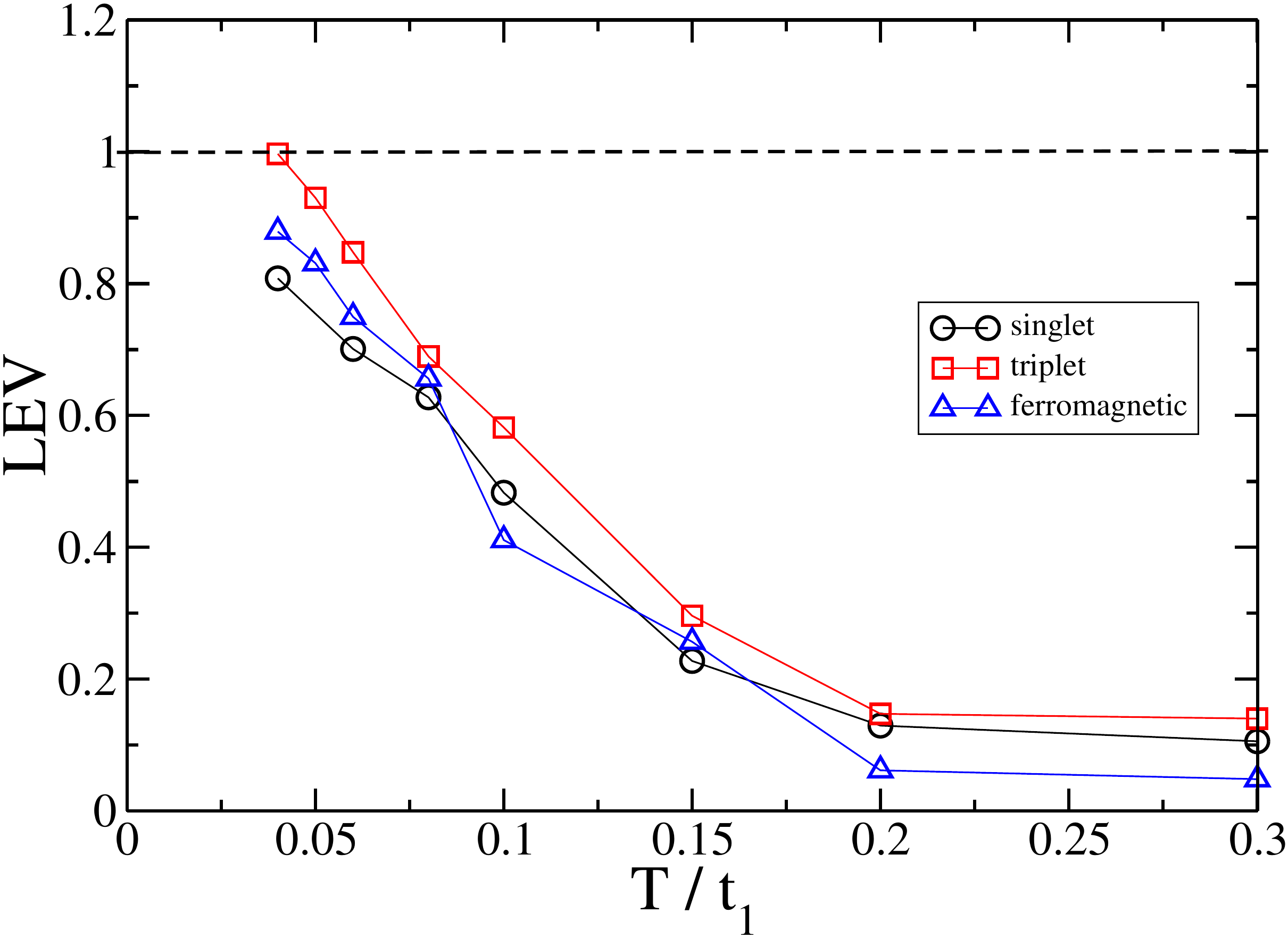}}
  \caption{(color online) LEVs close to the type-II vHS with $t_{2}=-0.5t_{1}$, $t_{3}=0.1t_{1}$, $n=0.4$ and $U=4t_{1}$. As temperature goes down, the ferromagnetic instability becomes the sub-leading below the leading two-fold degenerate, triplet pairing.}
  \label{fig:single_site_U4}
\end{figure}

Fig.~\ref{fig:single_site_U2} shows the LEVs obtained from DMFT+Parquet calculations, with parameter set: $t_{2}=-0.5t_{1}$, $t_{3}=0.1t_{1}$, $U=2t_{1}$ and band filling $n=0.4$ close to $n_{V}$. As temperature goes down, a triplet pairing becomes favored. Such triplet pairing has two-fold degenerate LEVs, and the corresponding leading eigenvectors have $p'_{x}=-p_{x}-p_{y}$ and $p'_{y}=-p_{x}+p_{y}$ symmetries, as shown in the insets of Fig.~\ref{fig:single_site_U2}. This implies that the triplet superconducting order parameter is doubly degenerate with components $p'_{x}$ and $p'_{y}$. In principle any linear combination of both $p$-wave components is possible below $T_c$. However, a $p+ip$ triplet pairing state is favored due to its largest condensation energy~\cite{Sigrist91,cheng2010stable}. Furthermore, since the system is spin-full, either chiral or time-reversal invariant $Z_2$ pairing could occur. Therefore, our findings support a possible odd-parity, topological superconducting phase in the vicinity of the type-II vHS, induced by repulsive interaction.

Interestingly, when the interaction strength increase, we find the two-fold degenerate $p$-wave pairing is still the leading instability. Fig.~\ref{fig:single_site_U4} show the same LEVs analysis from DMFT+Parquet but with a slightly larger $U=4t_{1}$. The two-fold degenerate triplet pairing is still the leading instability of the system, and the ferromagnetic instability wins over the singlet pairing to become sub-leading. In principle, since there is no obvious nesting in the FS for the parameter set we chose, triplet pairing and ferromagnetic instabilities will compete to be the leading one as the interaction strength further increases. 

\subsection{Dynamical cluster approximation}
Besides RPA and DMFT+Parquet, we also apply the DCA simulations~\cite{Hettler98, Hettler00} with interaction-expansion CTQMC~\cite{Rubtsov05} as cluster solver, to study the superconductivity instabilities. The DCA is a cluster DMFT method where one maps the lattice of the original system onto a periodic cluster of size $N_c=L_c^D$ ($D$ is the dimensionality) embedded in a self-consistently determined bath. Like the DMFT, the dynamical (frequency) correlations of the system are captured exactly, and the spatial short-ranged correlations (up to $L_c$) are treated explicitly while the long-ranged correlations are taken into account at mean-field level, the single and two-particle quantities computed from DCA have the momentum resolution up to $1/L_c$. DCA can be viewed as a step further of the DMFT+Parquet, once the convergence between cluster and bath is achieved, one can directly compute the two-particle vertex/correlation functions on the cluster, these two-particle quantities acquire exact frequency- and momentum-dependence (upto cluster size).

Although the DCA vertex/correlation functions are more accurate, one needs to be aware that the minus-sign problem is much severe in the cluster DCA simulation than that in the single-site DMFT simulation, especially for the high doping situation close to the vH filling $n_{V}$ and at strong interaction strength. The accessible temperature and interaction range for DCA simulation is much more restricted than those in the DMFT+Parquet, hence one needs to analyze the results from these two methods in a complementary manner.

\begin{figure}[tp!]
\centerline{
  \includegraphics[width=3.5in]{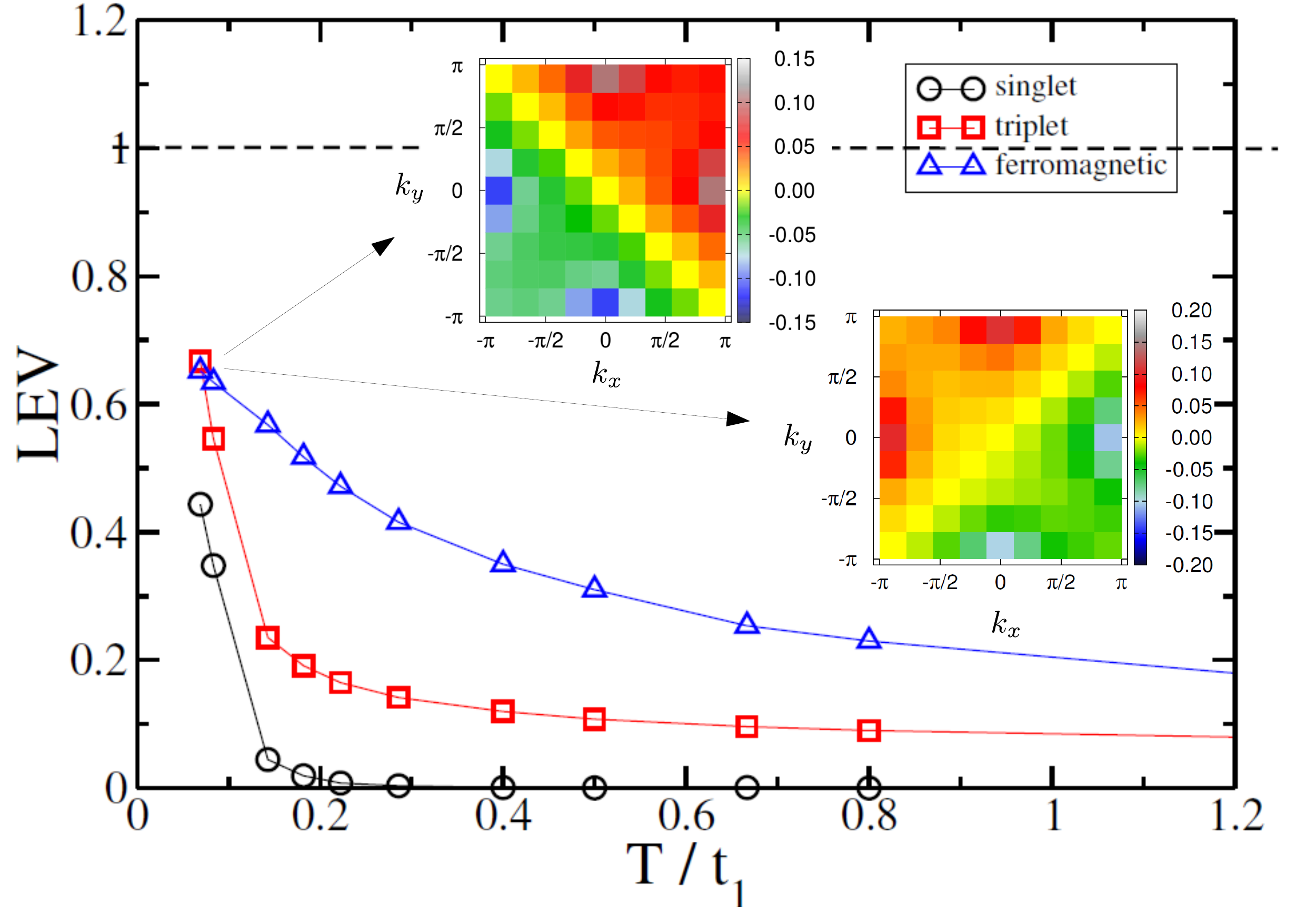}}
  \caption{(color online) LEVs close to the type-II vHS with $U=2t_{1}$, $t_{2}=-0.5t_{1}$, $t_{3}=0.1t_{1}$ and filling $n=0.4$. Results are obtained from 16-site DCA simulations. As temperature goes down, triplet pairing wins over the singlet pairing and ferromagnetic instabilities to become the leading instability. The corresponding leading eigenvectors are two-fold degenerate, one has $p'_{x}=p_{x}+p_{y}$ symmetry (upper inset) and the other has $p'_{y}=-p_{x}+p_{y}$ symmetry (lower inset).}
  \label{fig:cluster}
\end{figure}

Here the cluster size is $N_c=16$. As shown in the Fig.~\ref{fig:dispersion} (c), the $N_c=16$ cluster is able to capture the vH momenta $\mathbf{K}=(\pm\pi/2,0)$ and $\mathbf{K}=(0,\pm\pi/2)$. Hence the irreducible vertex function $\Gamma^{s/t}_{pp}(P,P',Q)$ is able to represent the pairing instabilities related to the type-II vHS. To avoid the serious minus sign problem in the large doping region of the Hubbard model, we study the doping $n=0.4$, slightly away from $n_{V}$. The interaction strength is set as $U=2t_{1}$ for comparison purpose. We obtain results with temperatures as low as $T=0.07t_{1}$.

We measure the two-particle correlation function in pp pairing channels, $\chi^{s/t}_{pp}(P,P',Q)$, where $P\equiv(\mathbf{K},\omega)$ and $Q\equiv(\mathbf{Q},\nu)$ with $\mathbf{K}$ and $\mathbf{Q}$ the momentum point on the cluster. We also use the Bethe-Salpeter equation to extract the irreducible vertex function, $\Gamma^{s/t}_{ph}(P,P',Q)$, and then construct the pairing matrix $\Gamma^{s/t}_{pp}(P,P',Q)\chi^{pp}_{0}(P',Q)$, where the particle-particle bare bubble $\chi^{pp}_{0}(P,Q)=G(-P)G(P+Q)$ with $G(P)$ the fully dressed single-particle Green's function, coarse-grained from lattice to cluster. We then perform the same analysis following Eq.~\ref{eq:pairing_eigen_equation}, to obtain the leading eigenvalues and eigenvectors of the pairing matrix.

The results are shown in Fig.~\ref{fig:cluster}. One can see that although the momentum-resolution is low ($K$, $K'$ only have 16 points), as temperature goes down, the triplet pairing LEVs (also two-fold degenerate) dominate over the singlet pairing LEVs, and there is a clear trend that the triplet LEVs will diverge as temperature becomes even lower. The leading eigenvectors of the two triplet pairing LEVs, again, have the the $p'_{x}=p_{x}+p_{y}$ and $p'_{y}=-p_{x}+p_{y}$ symmetries, as shown in the insets. At high temperature, the ferromagnetic LEV is larger, but it is clear from the temperature dependence that the triplet LEVs have a diverging trend at low temperatures, whereas the ferromagnetic LEV  only increases slowly. In fact at the lowest temperature we can access, $T=0.07t_1$, the triplet LEVs have already surpassed the ferromagnetic LEV.


\section{Concluding remarks}

Combining three different methods, RPA, large-scale DMFT+Parquet simulations, and large-scale DCA simulations, we have investigated the superconductivity instabilities and pairing symmetries in Hubbard model on square lattice whose band structure featuring type-II vHS. Close to the type-II vHS, we find the system process a doubly degenerate, odd-parity, $p$-wave triplet pairing state, triggered by the enhanced ferromagnetic fluctuations. From weak to relatively strong $U$, our findings provide evidence for a odd-parity, topological (either chiral or time-reversal invariant) $p+ip$ triplet superconducting phase in this model.

Our study is potentially relevant to quasi-2D superconducting materials whose band structure feature type-II vHS. One example is the recently discovered superconductor LaO$_{1-x}$F$_x$BiS$_2$\cite{mizuguchi20122,deguchi2013evolution}. It has layered structure, and the $p_x$ and $p_y$ orbitals of Bi (constitute most of the FS) form a 2D square lattice. This material can be tuned to type-II vHS by varying doping $x$ to around $x=0.5$\cite{Yao13,usui2012minimal} and features a superconducting dome\cite{deguchi2013evolution} with optimized $T_c$ around the VH filling. Functional renormalization group based study on the repulsive Hubbard-model representing the system yields odd-parity pairing symmetry\cite{Yang2013}, which is consistent with the insight gain here. Note that in real material, due to other different types of electron interactions~\cite{Liang2014}, the pairing symmetry can be different.  For another example, in the hexagonal systems, band-structure calculation~\cite{Chen14} shows in the doped BC$_{3}$ -- a graphenelike one-atom thick material -- the type-II vHS can occur at approximately 1/8 electrons per site. In the presence of repulsive interaction, combined renormalization group and RPA studies reveal time-reversal-invariant $p+ip$ topological pairing states in the system. Yet another example is the doped transition metal dichalcogenides Ir$_{1-x}$Pt$_{x}$Te$_{2}$~\cite{Kiswandhi2013,Fang2013,Qian2014}, Pt doped IrTe$_{2}$ becomes superconductor, while IrTe$_{2}$ itself has a structural transition. Experimentally it is shown that the structure transition is intimately associated with the type-II vHS originates from the Te $p_{x}+p_{y}$ orbitals~\cite{Qian2014}. It might well be the type-II vHS in this system is also related to the superconductivity in the doped case.

The results presented in this paper provide a general insight, that in the presence of repulsive interaction, in a finite doping range near the type-II vHS, odd-parity triplet superconductivity would be favored. We demonstrate such a insight by employing state-of-art large-scale quantum many-body numerical techniques on a simple and typical one-band model. As for a future direction beyond current model-level scheme, one needs go to real materials, such as the three systems mentioned above. There, more realistic multi-orbital models are appropriate. In such systems, it is possible to have both type-I and type-II vHS, various pairing channels associated with enven and odd parity will start to compete, it is then interesting to study orbital-selective superconducting instabilities, especially in the presence of spin-orbit coupling.

\begin{acknowledgments}
We would like to acknowledge Y. L. Wang, H. Li, X. Dai, G. Chen and K. Sun for helpful discussions. The DMFT+CTQMC simulations code belongs to the open source interacting quantum impurity solver toolkit -- the \textit{i}QIST package~\cite{iQIST2014}. This work is supported in part by the NSERC, CIFAR, and Centre for Quantum Materials at the University of Toronto (ZYM and HYK), the National Thousand-Young-Talents Program of China (ZYM and HY), and the NSFC under Grant No.11274041, 11334012 and the NCET program under Grant No. NCET-12-0038 (FY). Computations were performed on the GPC supercomputer at the SciNet HPC Consortium as well as at the National Supercomputer Center in Tianjin on the platform TianHe-1A.
\end{acknowledgments}

\bibliography{TypeIIVHS_Superconductivity_main_Aug15}

\end{document}